\documentclass[aip,jcp, preprint]{revtex4-1}
\usepackage{graphicx}
\usepackage{amsmath}
\usepackage{color}
\usepackage{threeparttable}
\begin{document}

\def\n{n}
\def\beq{\begin{equation}}
\def\eeq{\end{equation}}
\def\sss{\scriptscriptstyle\rm}
\def\x{_{\sss X}}
\def\c{_{\sss C}}
\def\s{_{\sss S}}
\def\xc{_{\sss XC}}
\def\dc{_{\sss DC}}
\def\ext{_{\rm ext}}
\def\ee{_{\rm ee}}ㅅ
\def\sint{ {\int d^3 r \,}}
\def\br{{\mathbf r}}

\title{Avoiding Unbound Anions in Density Functional Calculations}
\author{Min-Cheol Kim}
\affiliation{Department of Chemistry and Institute of Nano-Bio Molecular Assemblies, 
Yonsei University, 262 Seongsanno Seodaemun-gu, Seoul 120-749 Korea}
\author{Eunji Sim$^*$}
\affiliation{Department of Chemistry and Institute of Nano-Bio Molecular Assemblies, 
Yonsei University, 262 Seongsanno Seodaemun-gu, Seoul 120-749 Korea}
\author{Kieron Burke$^*$}
\affiliation{Department of Chemistry, University of California, Irvine, CA, 92697, USA}
\date{\today}

\begin{abstract}
Converged approximate density functional calculations usually do not bind anions, due to large self-interaction error. But Hartree-Fock calculations have no such problem, producing negative HOMO energies. A recently proposed scheme for calculating DFT energies on HF densities is shown to work very well for molecules, better than the common practice of restricting the basis set, except for cases like CN, where the HF density is too inaccurate due to spin contamination.
\end{abstract}

\maketitle

Anions and radicals are important for many applications, including
environmental chemistry,\cite{VMA95, HDT08, VEW10} semiconductors,\cite{TAX98, BAR10}
fullerene chemistry,\cite{CEP91, DC02, SL06, WWH06, WWY07} charge transfer,\cite{HZ98} and solar cells.\cite {KLC07, SCH10}
 Recently, electron affinities of biological species become of great interest, especially in studies of low-energy electron DNA damage.\cite{BCH00, LCS02, DLS09, GXS10, GXS10b, CGC10, KS10} Low-energy electrons cause single-strand breaking, double-strand breaking, and supercoil loss in DNA even below the DNA ionization potential. The electron affinity of DNA bases and base-pairs is important in determining damage mechanism.

Density functional theory (DFT) has become a standard method for electronic structure calculations in chemistry, and the standard functionals can be applied with standard basis sets to calculate electron affinities. The results are excellent, with mean absolute errors (MAE)
below about 0.2 eV.\cite{RTS02} However, there is a theoretical fly in the computational ointment: Inspection of the orbital energies show that the HOMO of the anion is usually positive. This implies that, in principle, the calculation is unconverged.\cite{RT97} 
If a sufficiently large basis set had been used, a fraction of the additional electron would ionize\cite{J10} and the HOMO drop to zero. 
This is due to the self-interaction error that all the standard density functional approximations suffer from. 
This error is especially large for anions, because of their additional electron. Self-interaction error produces an exchange-correlation potential that incorrectly decays exponentially in the asymptotic region, instead of decaying as $-1/r$. For atomic anions, a large positive barrier appears in the Kohn-Sham potential (See Fig. 1 of Ref \onlinecite{LFB10}) resulting in positive HOMO resonances. These metastable states are occupied and artificially bound by moderate basis sets (MBS), and so produce a positive HOMO. Because the positive barrer is often very wide, the total energy appears converged unless extreme basis functions are used.

There have been strong discussions about this issue.\cite{RT97, GS96, JD99, LFB10, LB10}
Users find reasonable results with MBS for most cases, and ignore the postive HOMO. Purists regard all such calculations as unconverged, and so their results are suspect.\cite{RT97} Pragmatists will report results with the standard methods, but attach a {\em caveat emptor} footnote.\cite{RTS02} The paradox has recently been addressed in several papers,\cite{LFB10, LB10} which explain how accurate results can come from such unconverged calculations, but also suggest an alternative procedure that avoids the dilemma: Evaluate the density functional total energies on Hartree-Fock (HF) densities. We refer to this method as HF-DFT.

Because HF is exact for one-electron systems, it has no self-interaction error, and its HOMOs are bound, even for anions. 
Electron affinity calculations for atoms and their anions show excellent results with either method,\cite{LFB10, LB10} with MAEs about 0.1 eV, about half of that for ionization potentials.

\begin{figure}[h]
\begin{center}
\includegraphics[width=3in]{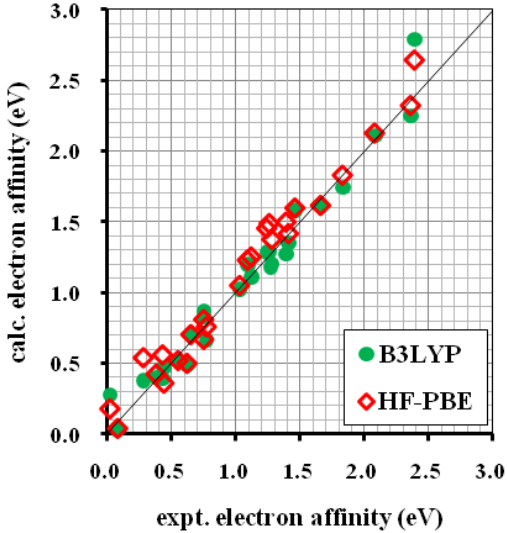}
\caption{Calculated electron affinities of molecules in the G2-1 set (excluding CN) and atoms of the first two rows of periodic table plotted against experimental electron affinity. B3LYP energies were evaluated on self-consistent densities, while PBE energies were evaluated on HF densities within the AVTZ basis set.}
\label{ea_cal_exp_plot}
\end{center}
\end{figure}

In the present work, we test the new procedure for the adiabatic electronic affinity of the molecules in the G2-1 data set.\cite{CRT91} All calculations are performed with TURBOMOLE 6.2.\cite{turbomole} The functionals used in DFT calculations are PBE,\cite{PBE96} B3LYP,\cite{B93, LYP88, b3lyp} and PBE0.\cite{PEB96} We use Dunning's augmented correlation-consistent pVTZ (AVTZ) basis set.\cite{D89, WD93} Structures of neutral molecules and molecular ions are optimized with self-consistent Kohn Sham-DFT and non-scaled zero-point vibrational correction with the same functional is added. For HF-DFT calculations, unrestricted HF calculations are performed on the DFT optimized structures. Based on these HF orbitals, the same functional used in structure optimization was selected for energy evaluation. We distinguish such calculations by HF-XC, where XC indicates the exchange-correlation approximation used. For all our calculations, SCF convergence was achieved with 10$^{-8}$ eV or lower energy difference and with 10$^{-8}$ root-mean-square density matrix element deviation. We exclude CN because the unrestricted HF solution of neutral CN is strongly spin contaminated.\cite {advchemphys} This results in both a large energy destabilization and also poor HF densities.\cite{SP73, supp1}

\begin{table}[h]
\caption{Electron affinities (EA) of molecules and HOMO eigen values of anions in the G2-1 set excluding CN (eV). All calculations were with the AVTZ basis set and using DFT optimized geometries. HF HOMO eigenvalues were evaluated from B3LYP geometries.}
\begin{center}
\begin{threeparttable}
\begin{tabular}{c| c |c c |c c | c c}
\hline\hline
& EA & \multicolumn{2}{c}{$\Delta$EA(MBS)} \vline& \multicolumn{2}{c}{$\Delta$EA(HF-DFT)}\vline & \multicolumn{2}{c}{$-\epsilon_{\sss HOMO}$}\\
\raisebox{2ex}{Mol.} & expt. & PBE & B3LYP & PBE & B3LYP & B3LYP & HF\\
\hline
CH       & 1.24 & 0.29 & 0.04 & 0.22 & -0.03 & -1.3 & 2.1 \\
CH$_2$   & 0.65 & 0.13 & 0.06 & 0.05 & 0.02 & -1.3 & 1.3 \\
CH$_3$   & 0.08 & 0.00 & -0.05 & -0.04 & -0.09 & -1.7 & 0.6 \\
NH       & 0.38 & 0.17 & 0.02 & 0.04 & -0.07 & -2.1 & 0.1 \\
NH$_2$   & 0.77 & 0.06 & -0.11 & -0.01 & -0.15 & -1.7 & 1.3 \\
OH       & 1.83 & 0.12 & -0.09 & 0.00 & -0.16 & -1.1 & 3.0 \\
SiH      & 1.28 & 0.12 & -0.08 & 0.10 & -0.12 & -0.8 & 1.5 \\
SiH$_2$  & 1.12 & 0.17 & -0.01 & 0.13 & -0.06 & -1.0 & 1.3 \\
SiH$_3$  & 1.41 & 0.01 & -0.06 & 0.01 & -0.05 & -0.3 & 1.8 \\
PH       & 1.03 & 0.04 & -0.01 & 0.02 & -0.01 & -1.1 & 0.9 \\
PH$_2$   & 1.27 & -0.02 & -0.09 & -0.01 & -0.08 & -1.0 & 1.2 \\
HS       & 2.36 & -0.03 & -0.11 & -0.04 & -0.11 & -0.2 & 2.6 \\
O$_2$    & 0.44 & 0.00 & 0.03 & -0.08 & -0.02 & -2.2 & 2.4 \\
NO       & 0.02 & 0.27 & 0.26 & 0.16 & 0.14 & -2.3 & 2.5 \\
PO       & 1.09 & 0.18 & 0.11 & 0.14 & 0.07 & -1.1 & 2.0 \\
S$_2$    & 1.66 & -0.07 & -0.04 & -0.04 & -0.02 & -0.5 & 2.2 \\
Cl$_2$   & 2.39 & 0.27 & 0.40 & 0.26 & 0.38 & 1.9 & 4.7 \\
\hline
MAE      & 0.00 & 0.11 & 0.09 & 0.08 & 0.09 & 2.2\tnote{*} & 0.8\tnote{*} \\
ME       & 0.00 & 0.10 & 0.02 & 0.05 & -0.02 & -2.2\tnote{*} & 0.8\tnote{*} \\
\hline\hline
\end{tabular}
\begin{tablenotes}
\item[*]Based on Koopman's theorem, mean absolute errors and mean errors are obtained by comparing the differences between $-\epsilon_{\sss HOMO}$ and EA.
\end{tablenotes}
\end{threeparttable}
\end{center}
\label{dEAtable}
\end{table}

In Fig. \ref{ea_cal_exp_plot}, we plot calculated versus experimental electron affinities, showing just how good the overall agreement is. In Table \ref{dEAtable}, we report results for both methods for all molecules in the G2-1 set, but with averages excluding CN. Averages are reported in MAE and mean of errors (ME). We find once again excellent results for almost all molecules, with either conventional DFT with MBS or the HF-DFT method, with MAEs again about 0.1 eV.	We find HF-PBE yields the best results overall, but differences are slight and not significant.

\begin{figure}[h]
\begin{center}
\includegraphics[width=3.5in]{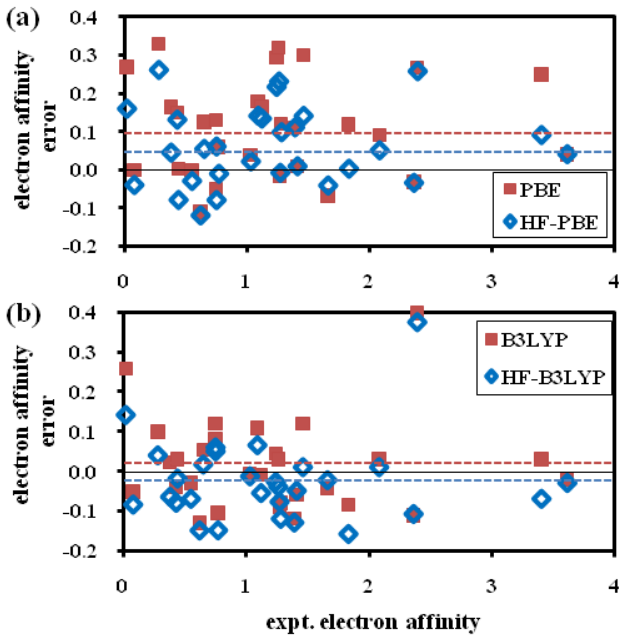}
\caption{Comparison of errors in electron affinities (eV) of molecules in the G2-1 molecule set excluding CN and atoms of the first two rows of periodic table. The colored dotted lines  indicate the mean error of each method. All calculations use AVTZ basis set.}
\label{ea_error_plot}
\end{center}
\end{figure}

In Fig. \ref{ea_error_plot}, we plot errors versus electron affinities, including also the results from atoms of Refs \onlinecite{LFB10, LB10}. We have excluded CN, where spin contamination of the HF wavefunction makes the HF density of the neutral too poor for accurate calculation. Spin contamination in HF for CN is $>50\%$, whereas no other molecule in the set has greater than 10\%. It is suggested that calculations with spin contamination higher than 10\% are not reliable.\cite{compchem} Notice that switching the density from self-consistent to HF always either reduces the electron affinity (sometimes increasing the error), or increases it by no more than 0.02 eV. In the case of PBE, the self-consistent electron affinities are mostly too large, a systematic error inherited from the local density approximation (LDA, sometimes called VWN\cite{VWN80}). This is reflected in the fact that the ME $\sim$ MAE on the scale of the MAE in Table \ref{dEAtable}. The reduction in electron affinities on using the HF density, which leads to subsequent reduction in MAE and large reduction in ME, shows that this is largely a self-interaction error in the density, not the energy functional. On the other hand, B3LYP is a hybrid functional with empirical parameters. It already cancels some self-interaction error, and has smaller MAE. But the ME is much smaller than the MAE, showing that its errors have random signs, i.e., much less systematic than those of PBE. Inserting the HF density overcounts the self-interaction,
does not improve MAE, and even worsens ME. 
To check our interpretation of the effect of the hybrid,
we applied another functional, the non-empirical hybrid PBE0\cite{PEB96}, and found results with the same trends,
but higher MAEs (1.4 eV with or without HF densities).\cite{supp2}
We also show the HOMO energies for the anions, in both HF and self-consistent calculations. All species except Cl$_{2}$ have positive HOMO in the DFT calculations, indicating their unbound nature. A sufficiently large basis\cite{JD99} would reduce this value, but this effect may not be noticeable with any standard basis set. On the other hand, although the HF HOMO's are negative, they are not an accurate guide to the true electron affinities. Using Koopman's theorem, one may estimate electron affinity from the HOMO energy of anion, but both relaxation and correlation effects are so large that Koopman's theorem is unhelpful here. Since the HF density is calculated upon geometries optimized from DFT, the resulting HF-DFT energy will not typically be a minimum in the HF-DFT potential energy surface. This shows further improvement may be made in HF-DFT by development of potential energy surface scan and optimization techniques.

\begin{figure}[h]
\begin{center}
\includegraphics[width=3.5in]{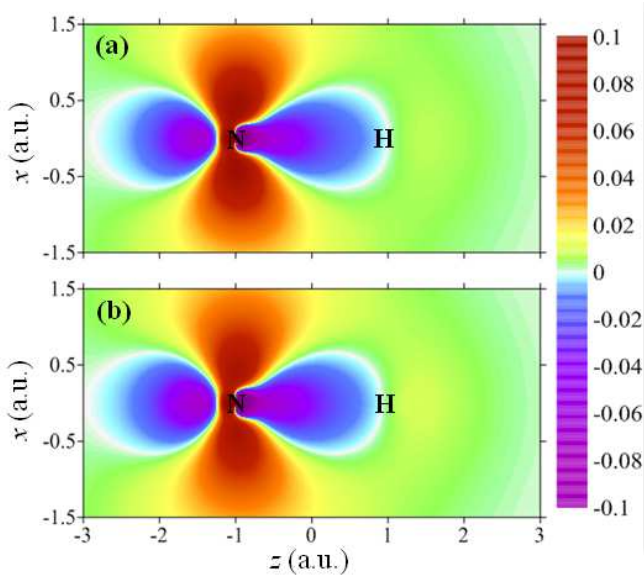}
\caption{Cross sections of electron affinity densities (anion - neutral) along the molecular axis of (a) PBE and (b) HF plotted for NH.}
\label{EA_2D_density}
\end{center}
\end{figure}

\begin{figure}[h]
\begin{center}
\includegraphics[width=3.5in]{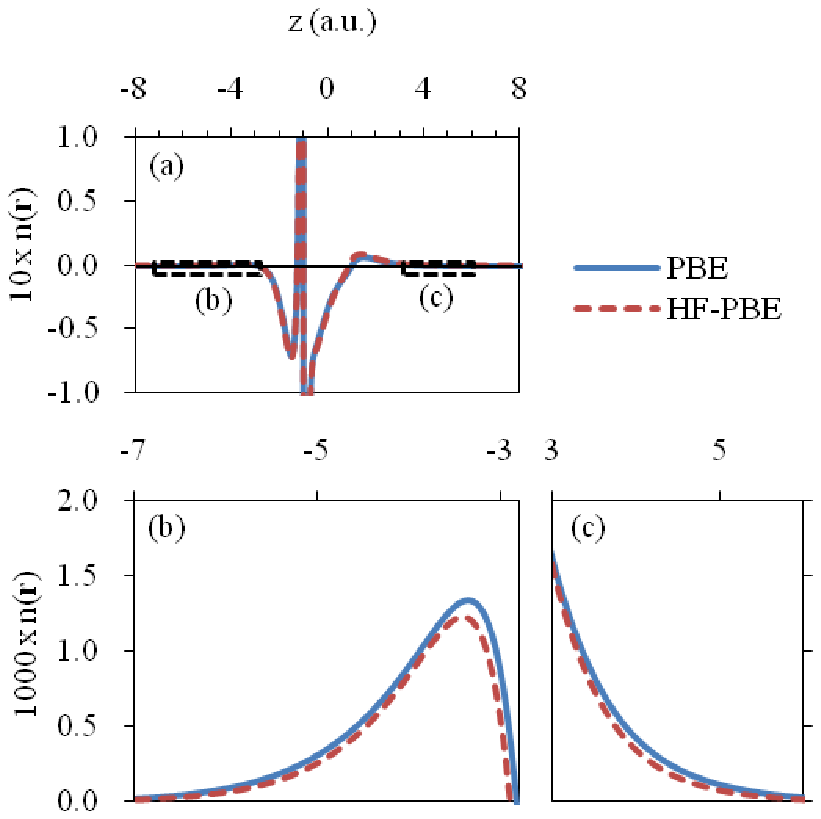}
\caption{Local electron affinity densities (anion - neutral) along the molecular axis ($z$-axis)  in Fig. \ref{EA_2D_density}. N is positioned at $z = -0.99$ and H is at $z = 0.99$. Regions in (a) are magnified into (b) and (c) for clarity.}
\label{EA_line_density}
\end{center}
\end{figure}

A useful tool for understanding these effects is the electron affinity density:
\beq
\n_{\scriptstyle\rm EA}(r) = \n_{-1}(r) - \n_{0}(r)
\eeq
where $n_{0}(r)$ is the charge density of the neutral, and $n_{-1}(r)$ is that of the anion. Fig. 6 of Ref \onlinecite{LB10} shows this for the Cl atom and anion, and how the HF electron affinity density is more compressed than that of using MBS and standard functionals. We plot the cross sections of electron affinity densities of NH along the molecular axis for different methods. In Fig. \ref{EA_2D_density}, the electron affinity densities of self-consistent and HF calculations are plotted for NH. NH, which has the largest deviation between the electron affinity error of PBE and HF-PBE, the electron affinity density of the two is nearly identical. Nonetheless, the self-consistent density is more diffuse than the HF density due to the electron leakage in the anion as shown in Fig. \ref{EA_line_density}. To confirm this is not an artificial effect from the geometry difference in neutral and anion, we present the local vertical electron detachment densities, which are electron affinity densities using the same geometry for the anion and neutral.\cite{supp3}

In summary, we have shown that the methods proposed in Refs \onlinecite{LFB10, LB10} work just as well for the small molecules of the G2-1 data set, so long as the HF densities are close to the true densities. The results are equally good with HF-PBE as with B3LYP, perhaps slightly better. On the other hand, we found no case where the limited basis set approach fails. Presumably, the unphysical barrier holding the additional electron in for atomic anions \cite{RT97} is sufficiently large for our molecules that standard basis sets show no sign (other than a positive HOMO) that the state being calculated is a resonance, rather than an eigenstate. We recommend that the HF-DFT method be applied more broadly for electron affinity calculations, especially for cases where DFT with MBS is believed to be inaccurate.

\section*{Acknowledgement}
We thank Prof. Fillip Furche and Dr. Donghyung Lee for fruitful discussions. This work was supported by the global research network grant funded by the Korean Government (NRF-2010-220-C00017) and by the national research foundation (NRF-2010-0016487, NRF-2010-0017172) in which part of calculations was performed by using the supercomputing resource of the Korea Institute of Science and Technology Information (KISTI). M-C thanks the fellowship of the BK 21 program from MOEHRD.

\vspace{0.5cm}\rule{\linewidth}{0.5mm}\vspace{0.2cm}
* Corresponding Author: esim@yonsei.ac.kr, kieron@uci.edu \vspace{0.5mm}


\begin{thebibliography}{43}
\bibitem [{b3l()}]{b3lyp}%
  \BibitemOpen
  \href@noop {} {}\bibinfo {note} {{We used the {V}{W}{N}-{V} functional as the
  local correlation functional, which is "functional {V}" in ref
  \onlinecite{VWN80}.}}\BibitemShut {Stop}%
\end{thebibliography}
\end{document}